%% file: CUW10draft.tex
\documentclass[12pt]{article}
\usepackage[latin1]{inputenc}
\usepackage{graphics, bm,amsmath, amsfonts, amssymb,  amstext}
\usepackage{amsthm,wrapfig, amsopn, latexsym,  epsfig,multicol}
\usepackage[colorlinks, citecolor=red,linkcolor=blue]{hyperref}

\title{PARAMETER ESTIMATION FOR FRACTIONAL POISSON PROCESSES}

\author{Dexter O. Cahoy$^{\S}$ \\\and Vladimir V. Uchaikin$^{\dag}$ \\ \and  Wojbor A. Woyczynski$^{\ddag}$ 
}

\begin{document}
\date{}
\maketitle
\renewcommand{\thefootnote}{\fnsymbol{footnote}}
\footnote[4]{Department  of Mathematics and Statistics, Louisiana Tech University, USA}
\footnote[2]{ Dep't of Statistics, Case Western Reserve University, USA, email: {\tt waw@case.edu}}
\footnote[3]{ Dep't of Theoretical and Mathematical physics, Ul'yanovsk State University, Russia}


\setlength{\parskip}{1ex plus 0.5ex minus 0.2ex}
\setlength{\parindent}{10mm}
\input{abst2.tex}

\section{Introduction}

The paper proposes a formal  estimation procedure  for  parameters of the {\it fractional Poisson process}  (fPp). Such procedures are needed to make the fPp model  usable  in applied situations. Different versions of fPp have  been studied recently  by several authors, see, in particular, Repin and Saichev (2000), Wang and Wen (2003), Wang, Wen and Zhang (2006) and Laskin (2003), so we start our exposition from the basic definitions to make it clear  which stochastic model we are working with. Some of the preliminary results on this model appeared in Uchaikin, Cahoy and Sibatov (2008) but we restate them in the first couple of sections for the sake of completeness of presentation. The basic idea of fPp, motivated by experimental data with long memory (such as some network traffic, neuronal firings, and other signals generated by complex systems), is to make the standard Poisson model more flexible by permitting non-exponential, heavy-tailed  distributions of interarrival times. However,  the price one has to pay for such flexibility is loss of the Markov property, a similar situation to that encountered in the case of certain anomalous diffusions, see, e.g.,  Priyatinska, Saichev and Woyczynski (2005).  To partly replace this loss  one demands some scaling properties of the interarrival times' distributions which makes other tools available; in this paper they are  the fractional calculus and
the link between fractional Poisson process   and $\alpha$-stable L\'evy densities. Based on the latter connection we  establish the  asymptotic normality of our estimators for  the two parameters  appearing  in our fPp model:   the  intensity rate $\mu$,  and the  fractional exponent $\nu$.   This fact permits construction of the  corresponding confidence intervals.   The properties of the estimators    are then tested using synthetic data.

The paper is composed as follows:  Section 2 introduces the basic definition of fPp and the fractional calculus tools needed to study it. Section 3 proves the basic structural theorem relating fPp to $\alpha$-stable L\'evy  random variables which makes efficient simulation of the former possible. In Section 4, we describe nontrivial scaling limits of the marginal distributions of fPp. Section 5 introduces the concept of the method-of-moments estimators in the fPp context and calculates them. They are proven asymptotically normal in Section 6. Finally, we test our procedures numerically on simulated data in Section 7. The concluding remarks in Section 8  are then followed by two brief appendices, one on  $\alpha^+$-stable stable densities, and one on an alternative fPp model.


\section{FPp interarrival time}

The fractional Poisson process $N_\nu(t),\, 0<\nu\le 1, \, t>0,$ was defined in Repin and Saichev (2000)  via the following formula for
 the Laplace transform of the p.d.f $\psi_\nu(t)$  of  its i.i.d. interarrival  times $T_i,i=1,2,\dots$:
\begin{equation}
\{\textsf{L}\psi_\nu(t)\}(\lambda)\equiv\widetilde{\psi}_\nu(\lambda)\equiv    \int\limits_0^\infty
e^{-\lambda t}\psi_\nu(t)dt =\frac{\mu}{\mu+\lambda^\nu},  \label{2e4c}
\end{equation}
where $\mu >0$ is a  parameter.
For $\nu=1$, the above transform  coincides with the Laplace transform
\[
\widetilde{\psi}_1(\lambda)=\frac{\mu}{\mu+\lambda}.
\]
of the exponential interarrival time density of
 the ordinary Poisson process with parameter $\mu= {\bf E}N_{1}(1)$.

Using the inverse Laplace transform the above cited authors  derived the  {\it singular integral equation}  for $\psi_\nu(t)$:
\[
\psi_{\nu}(t)+\frac{\mu}{\Gamma(\nu)}\int\limits_0^t\psi_\nu(\tau){ d\tau \over [\mu(t-\tau)]^{1-\nu }}=
\frac{\mu^\nu}{\Gamma(\nu)} t^{\nu-1},
\]
which is equivalent to the   {\it fractional differential equation},
\[
_0D_t^{\nu}\psi_{\nu}(t)+\mu\psi_\nu(t)=\delta (t),
\]
where the Liouville derivative   operator $_0D_t^{\nu}=d^\nu/dt^\nu$  (see, e.g., Kilbas, Srivastava and Trujillo (2006)) is  defined via the formula
\[
_0D_t^{\nu} \psi_\nu(\tau) =\frac{1}{\Gamma (1-\nu)}\frac{d}{dt}\int\limits_0^t \psi_\nu(\tau) \frac{d\tau}{[\mu(t-\tau)]^{1-\nu}}.
\]
These characterizations permitted them  to obtain the following integral representation for the p.d.f.  $\psi_\nu(t)$,
\begin{equation}
 \psi_\nu(t)=\frac{1}{t}\int\limits_0^\infty
e^{-x}\phi_\nu(\mu t/x)dx, \label{2e6}
\end{equation}
where
\[
\phi_\nu(\xi)=\frac{\sin(\nu\pi)}{\pi[\xi^\nu+\xi^{-\nu}+2\cos(\nu\pi)]},
\]
and demonstrate that the tail probability distribution of the waiting time $T$ is of the form
\begin{equation}
 {\bf P}(T>t)=\int_t^\infty \psi_\nu(\tau)\,d\tau=E_\nu(-\mu t^\nu), \label{2e5}
\end{equation}
where
\begin{equation} \label{MLfunction}
E_\nu(z)=\sum\limits_{n=0}^\infty\frac{z^n}{\Gamma(\nu
n+1)}
\end{equation}is the Mittag-Leffler   function  (see, e.g., Kilbas, Srivastava and Trujillo (2006)).
\bigskip

{\it Remark 2.1.} Observe that the Mittag-Leffler function is a fractional generalization
of the standard exponential function $\exp (z)$; indeed
  $ E_1(z)=\exp (z)$. It  has been widely used to describe probability distributions appearing  in finance and economics, anomalous diffusion,  transport of charge carriers in semiconductors,  and  light propagation through random media (see, e.g. Piryatinska, Saichev and Woyczynski (2005), and Uchaikin and Zolotarev (1999)).

  \bigskip

In view of (\ref{2e5}-\ref{MLfunction}), the interarrival  time density for the fractional Poisson process can be easily shown to be
\begin{equation}
\psi_\nu(t)=\mu t^{\nu-1} E_{\nu,\,\nu}(-\mu t^{\nu}), \qquad t\ge 0, \label{2e16}
\end{equation}
where
\[
E_{\alpha,\,\beta}(z) = \sum_{n=0}^{\infty}\frac{z^{n}}{\Gamma
(\alpha n + \beta)}
\]
is the generalized, two-parameter Mittag-Leffler function. Also, the above information automatically gives the p.d.f.
\begin{equation}
 f_n^\nu(t)= \mu^n \nu \frac{ t^{\nu n-1}}{(n-1)!}E_\nu^{(n)}\big(- \mu t^{\nu} \big),  \label{2e20}
\end{equation}
 of the $n$-the arrival time, $A_n=T-1+\dots+T_n$, because, obviously, its Laplace transform,   \[
\textsf{L} \big\lbrace f_n^\nu(t) \big\rbrace = \frac{\mu^n}{(\mu + \lambda^\nu)^{n}}.
\]
As $\nu\to 1$, the above distribution converges to the classical Erlang distribution.

\bigskip

{\it Example 2.1.} For some values of $\nu$, the p.d.f. of the interarrrival  times can be calculated more explicitly. In particular, consider
\[
\psi_{1/2}(t) =\mu t^{1/2-1}E_{1/2,1/2}\left( -\mu t^{1/2} \right), \qquad t\ge 0,
\]
where
\begin{equation}
E_{1/2,1/2}\left( -z \right)
=\sum_{n=0}^{\infty}\frac{(-z)^{n}}{\Gamma \left(\frac{n}{2} +
\frac{1}{2}\right)}
=\frac{1}{\sqrt{\pi}}-zE_{1/2,1}(-z).
\end{equation}
Using the identity,
\[
E_{1/2,1}(-z) =e^{z^2}\text{Erfc} (z),
\]
where
\[
\text{Erfc}(t)=\frac{2}{\sqrt{\pi}}\int _{z}^\infty e^{-u^2}du,
\]
 is the complementary error function, we  obtain,
\begin{align}
 \psi_{1/2} (t) & =\mu t^{-1/2} \left( \frac{1}{\sqrt{\pi}} - \mu t^{1/2} e^{\left( \mu t^{1/2}\right)^2}\text{Erfc} (\mu \sqrt{t})\right) \notag\\
 &= \frac{\mu}{\sqrt{\pi t}}-\mu^2e^{\mu^2t}\text{Erfc} (\mu
 \sqrt{t}), \qquad t\ge 0. \label{2e16a}
\end{align}


  \bigskip

In another approach to the study of $N_\nu(t)$,  Laskin (2003) used the fractional Kolmogorov-Feller-type differential equation system
\begin{equation}
_0D_t^{\nu}P_n^\nu(t)=\mu[P_{n-1}^\nu(t)-P_n^\nu(t)] +
\delta_{n0}\frac{t^{-\nu}}{\Gamma (1-\nu)}, \quad n=1,2,\dots, \label{2e10}
\end{equation}
 to characterize  the  1-D probability distributions $P^\nu_n(t)={\bf  P} \,(N_\nu(t)=n)$.  The solutions of the above system of
equations~(\ref{2e10})    can be calculated  to be
 \begin{equation}
G_\nu(u,t)\equiv {\bf E }u^{N_\nu(t)}=E_\nu\left(\mu t^\nu(u-1)\right). \label{2e11}
\end{equation}
Hence, expanding $G_\nu(u,t)$ over $u$,  and rearranging
(\ref{2e11})), we find
\begin{equation}
P_n^\nu(t)=\frac{(-z)^n}{n!}\frac{d^n}{dz^n}E_\nu (z) \bigg|_{z=-\mu
t^\nu}=\frac{(\mu
t^\nu)^{n}}{n!}\sum_{k=0}^{\infty}\frac{(k+n)!}{k!}\frac{(-\mu
t^{\nu})^k}{\Gamma (\nu (k+n) +1)}.\label{2e14}
\end{equation}
Equivalently,  one can show (see, Laskin (2003)) that  the moment generating function
(MGF) of the fractional Poisson process $N_\nu(t)$ is of the form
\begin{equation}
M_\nu(s,t) \equiv {\bf E}\,  e^{-sN_\nu(t)}= \sum_{m=0}^{\infty}\frac{\left[\mu t^\nu\left(e^{-s}-1 \right)\right]^m}{\Gamma ( \nu  m+1)},
\end{equation}
which permits calculation (see, Table 1) of the fPp's moments via the usual formula,
\[
\textsf{E} \left[N_\nu(t)\right]^k = \left( -1\right)^k\frac{\partial^k}{\partial s^k} M_\nu(s,t)\big|_{s=0}.
\]

\begin{table}[h!t!b!p!]
 \caption{\emph{Properties of fPp compared with those of the Poisson
process.
}} \label{t1}
\vskip 5mm
        \centerline {
\begin{tabular}{|c||c|c|}
\hline
& Poisson process $(\nu=1)$ & Fractional Poisson Process $(\nu < 1)$\\
\hline \hline
& & \\
$P_0(t)$& $e^{-\mu t}$ & $E_{\nu}(-\mu t^{\nu} )$\\
& & \\
$\psi (t)$&$ \mu e^{-\mu t} $& $\mu t^{\nu-1} E_{\nu,\,\nu}(-\mu t^{\nu})$\\
& & \\
$P_n(t)$& $\frac{(\mu t)^{n}}{n!}e^{-\mu t}$ & $\frac{(\mu t^\nu)^{n}}{n!}\sum_{k=0}^{\infty}\frac{(k+n)!}{k!}\frac{(-\mu t^{\nu})^k}{\Gamma (\nu (k+n) +1)}$\\
& &\\
$\mu_{N(t)}  $& $\mu t$ &$\frac{\mu t^{\nu}}{\Gamma (\nu + 1)}$\\
& &\\
$ \sigma^{2}_{N(t)}$& $\mu t$ & $\frac{\mu t^{\nu}}{\Gamma (\nu + 1)}\bigg\lbrace{1 + \frac{\mu t^{\nu}}{\Gamma (\nu + 1)}\left[\frac{\nu B(\nu, 1/2)}{2^{2\nu-1}}-1 \right] \bigg\rbrace},$\\
& &$B(\alpha, \beta) = \frac{\Gamma (\alpha) \Gamma (\beta)}{\Gamma (\alpha + \beta)}$\\
& &\\
$\textsf{E} \left[N(t)\right]^k  $&$ \frac{\partial^k}{\partial s^k} s^k \exp\left[ \mu(s-1)t\right]\big|_{s=0}$&$ \left( -1\right)^k\frac{\partial^k}{\partial s^k} \sum_{m=0}^{\infty}\frac{\left[\mu t^\nu\left(e^{-s}-1 \right)\right]^m}{\Gamma (m \nu +1)}\big|_{s=0}$\\
& & \\
\hline
\end{tabular}
}
\end{table}

More recently, Mainardi et al. (2004, 2005) provided an approach to  fPp based on  analysis of the  survival probability function $ \Theta (t)=P(T>t)$. They have    shown that   $ \Theta (t)$ satisfies the fractional differential equation
\begin{equation}
_0D_t^{*\nu}\Theta (t) = -\mu \Theta (t),\qquad t \geq 0,\; \qquad \Theta (0^+) = 1, \label{2e17}
\end{equation}
where
\[
_0D_t^{*\nu}f(t) = \left\{
                           \begin{array}{ll}
                             \frac{1}{\Gamma (1-\nu)}\int_0^t \frac{f^{(1)}(\tau)}{(t-\tau)^\nu}d\tau, & 0<\nu<1; \\
& \\
                             \frac{d}{dt}f(t), & \nu=1.
                           \end{array}
                         \right.
\]
is the so-called Caputo derivative. Obviously, for the standard Poisson process  with parameter $\mu$,   $\Theta (t)$ satisfies the ordinary differential equation
\[
\frac{d}{dt}\Theta (t) = -\mu \Theta (t),\qquad  t \geq 0,\; \qquad \Theta (0^+) = 1.
\]


Some characteristics of the classical and fractional Poisson processes are compared in Table 1, above.


\section{Simulation of fPp interarrival times $(T_i)$}

Simulation of the usual Poisson process is very easy and efficient because, given a random variable $U$, uniformly distributed on $[0,1]$,   the random variable  $ {|\ln U|}/{\mu}$ has  the exponential distribution with parameter $\mu$.  With the interarrival time for fPp exhibiting a  more complicated structure described in Section 2 the issue of an efficient simulation of fPp depends on finding a representation for the Mittag-Leffler function which is more computationally convenient than the series (\ref{MLfunction}). Here, the critical observation is that the interarrival times $T_i\stackrel{d}{=}T$, are equidistributed with the random variable,
$$
T' = \frac{|\ln  U|^{1/\nu}}{\mu^{1/\nu}}S_{\nu},
$$
where $S_{\nu}\ge 0$ is a completely asymmetric $\nu$-stable random variable (see, Appendix 1) with the p.d.f.  $g_{\nu}(s)$ possessing  the Laplace transform
\begin{equation}
\label{laplacestable}
\int_{0}^{\infty}  g_{\nu}(s) e^{-\lambda s}\,ds= \exp(-\lambda^{\nu}).
\end{equation}

The verification of the above statement is  straightforward  in view of (\ref{2e5}) and the   integral representation for the Mittag-Leffler function implied by (\ref{laplacestable}); cf., e.g., Uchaikin and Zolotarev (1999):
\begin{align}
{\bf P}(T'>t)&= {\bf P}\left (\frac{|\ln  U|^{1/\nu}}{\mu^{1/\nu}}S_{\nu}>t\right)\nonumber\\
&=\int_{0}^{1}{\bf P}\left ( S_{\nu}>{ t \mu^{1/\nu} \over (-\ln  (1-u))^{1/\nu} } \right)\,du\nonumber\\
&=\int_{0}^{\infty}{\bf P}\left ( S_{\nu}>{ t \over \tau^{1/\nu} } \right) \mu e^{-\mu\tau}\,d \tau\nonumber\\
&=\int_{0}^{\infty}\left(\int_ {t/\tau^{1/\nu}}^{\infty} g_{\nu}(s)\,ds\right)  \mu e^{-\mu\tau}\,d \tau\nonumber\\
&=\int_{0}^{\infty} \int_ {t^\nu/s^{\nu}}^{\infty} g_{\nu}(s)   \mu e^{-\mu\tau}\,d\tau\,d s\nonumber\\
&=\int_{0}^{\infty}   g_{\nu}(s)    e^{-\mu t^{\nu}/s^{\nu}} \,d s\nonumber
 =E_{\nu}(-\mu t^{\nu})={\bf P}(T>t).
\end{align}

Utilizing the well known  Kanter (-Chambers-Mallows) algorithm (see, e.g., Kanter (1975)) we obtain the following corollary providing an algorithm for simulation of the fPp interarrival times:

 \noindent \newtheorem*{cor1}{Corollary}
\begin{cor1}
Let  $U_1,\ U_2$, and $U_3$, be  independent,   and uniformly distributed in
$[0,1]$. Then the fPp interarrival time
\begin{equation}
T \stackrel{d}{=}
\frac{|\ln
U_{1}|^{1/\nu}}{\mu^{1/\nu}}\frac{\sin(\nu\pi U_2)[ \sin((1-\nu)\pi
U_2)]^{1/\nu-1}}{[\sin (\pi U_2)]^{1/\nu}|\ln
U_3|^{1/\nu-1}}.\label{e7}
\end{equation}

 \end{cor1}

\begin{figure}[h!t!b!p!]
\centering 
\includegraphics[width=.850\textwidth]{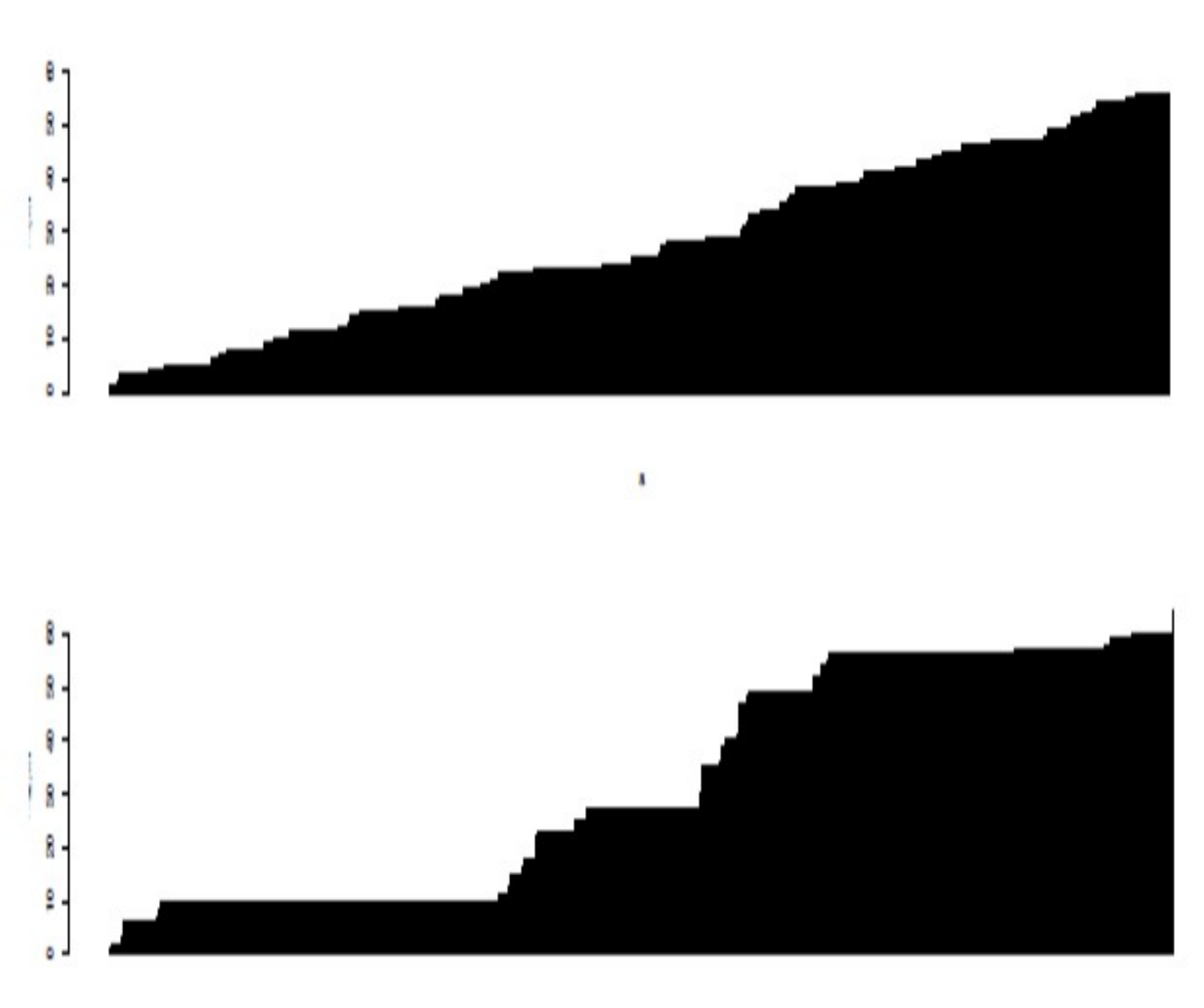}
 \caption{\emph{Sample trajectories of:  (a)  standard Poisson process, (b) fPp with parameter $\nu=1/2.$ }}
         \renewcommand\belowcaptionskip{0pt}
       \label{Figure1}
\end{figure}

 A comparison of sample trajectories for the standard Poisson process and an fPp, with parameter $\nu=1/2$, can be seen in Figure \ref{Figure1}.

\section{Scaling limit for  fractional Poisson distribution}

For the standard Poisson process,  $N(t)=N_{1}(t)$, the central limit theorem and infinite divisibility of the Poisson distribution give us immediately the following Gaussian scaling limit of distributions: as $\bar n=\mu t\to \infty$,
$$
{N(t)-\bar n \over
\sqrt{\bar n}}\stackrel{d}{\Longrightarrow} N(0,1)
.
$$
A more subtle,  skew-normal approximation to the Poisson distribution is provided by the following  formula: for $n=0,1,2,\dots,$
$$
{\bf P} (N(t)\le n)\approx \Phi (z)-{1 \over 6 \sqrt{\bar n}}(z^2-1)\phi(z),
$$
where $z=(n+{1 \over 2}-\bar n)/\sqrt{\bar n}$, and $\Phi$, and $\phi$,  are standard normal c.d.f, and p.d.f., respectively. The above formula, used to calculate the probabilities ${\bf  P} (m<N(t)\le n)$ (including ${\bf  P} (N(t)= n)$), guarantees, uniformly over $n,m$,  errors  not worse than $1/(20 \bar n)$ (as opposed to errors of the order $1 /\sqrt {\bar n}$ if the skewness correction term is dropped), see, e.g., Pitman (1993), p. 225.

 Considering the case of the  fPp,  $N_{\nu}(t)$, and introducing the standardized random variable
 $$
 Z_{\nu}= {N_{\nu}(t)\over \bar n_{\nu}},  \qquad {\rm where} \qquad  \bar n_{\nu}={\bf E}N_{\nu}(t)={\mu t^{\nu}\over \Gamma(\nu +1)}
 ,
 $$
    and substituting $u=e^{-\lambda/\bar n_{\nu}}$ in (10), we get
  the Laplace transform
 $$
 \textsf{E}e^{-\lambda
Z_{\nu}}=E_\nu(\bar n_{\nu}\Gamma(\nu+1)(e^{-\lambda/\bar n_{\nu}}-1)), \qquad \lambda>0,
$$
which has, for  large $ \bar n_{\nu} $ (i.e. large $t$) the asymptotics
$$
\textsf{E}e^{-\lambda Z_{\nu}} \sim E_\nu(-\lambda'),\quad
\lambda'=\lambda\Gamma(\nu+1).
$$
Since,
$$
E_\nu(-\lambda')=\nu^{-1}\int\limits_0^\infty\exp(-\lambda'x)g_{\nu} (x^{-1/\nu})x^{-1-1/\nu}dx
$$
$$
=\int\limits_0^\infty e^{-\lambda
z}\left\{\frac{[\Gamma(\nu+1)]^{1/\nu}}{\nu}g_\nu\left(\left(\frac{z}{\Gamma(\nu+1)}\right)^{-1/\nu}\right)
z^{-1-1/\nu}\right\}dz,
$$
where $g_{\nu}(s)$ is the $\nu$-stable p.d.f.,  see  Uchaikin and
Zolotarev (1999), formula (6.9.8),    the random variable $Z_{\nu}$ has,  for $\bar n_{\nu} \to \infty$,  a non-degenerate  limit
distribution  with the p.d.f.
\begin{equation}\label{f_nu_poisson}
f_\nu(z)=\left\{\frac{[\Gamma(\nu+1)]^{1/\nu}}{\nu}g_\nu\left(\left(\frac{z}{\Gamma(\nu+1)}\right)^{-1/\nu}\right)
z^{-1-1/\nu}\right\},
\end{equation}
with moments
$$
\langle
Z^k\rangle=\frac{[\Gamma(1+\nu)]^k\Gamma(1+k)}{\Gamma(1+k\nu)},
$$
see Uchaikin (1999).
Making use of the series expansion  for $g_\nu$, we obtain the series expansion
$$
f_\nu(z)= \sum_{k=0}^{\infty}\frac{(-z)^k}{k!\Gamma
(1-(k+1)\nu)[\Gamma (\nu +1)]^{k+1}}.
$$
Note that
$$
 f_\nu(0)= \frac{1}{\Gamma(1 + \nu)\Gamma(1-
\nu)}=\frac{\sin(\nu \pi)}{\nu \pi}. \nonumber
$$

It is also worth to note, that $\langle Z^0\rangle=1,\ \langle
Z^1\rangle=1$ and $\langle Z^2\rangle=2\nu{\rm B}(\nu, 1+\nu)$, so
that the limit relative fluctuations is given by
$$
\delta_\nu\equiv\sigma_{N(t)}/\langle N\rangle=\sqrt{2\nu{\rm
B}(\nu,
1+\nu)-1}=\begin{cases}1,\ \nu=0,\\
\sqrt{\pi/2}-1,\ \nu=1/2\\ 0,\ \nu=1.\end{cases}
$$

For $\nu=1/2$, one can obtain an explicit expression for
$f_\nu(z)$ :
$$
f_{1/2}(z)=\frac{2}{\pi}e^{-z^2/\pi},\ z\geq 0.
$$

The above family of limiting distributions is plotted below.

\begin{figure}[h!t!b!p!]
\centering
\includegraphics[width=.75\textwidth]{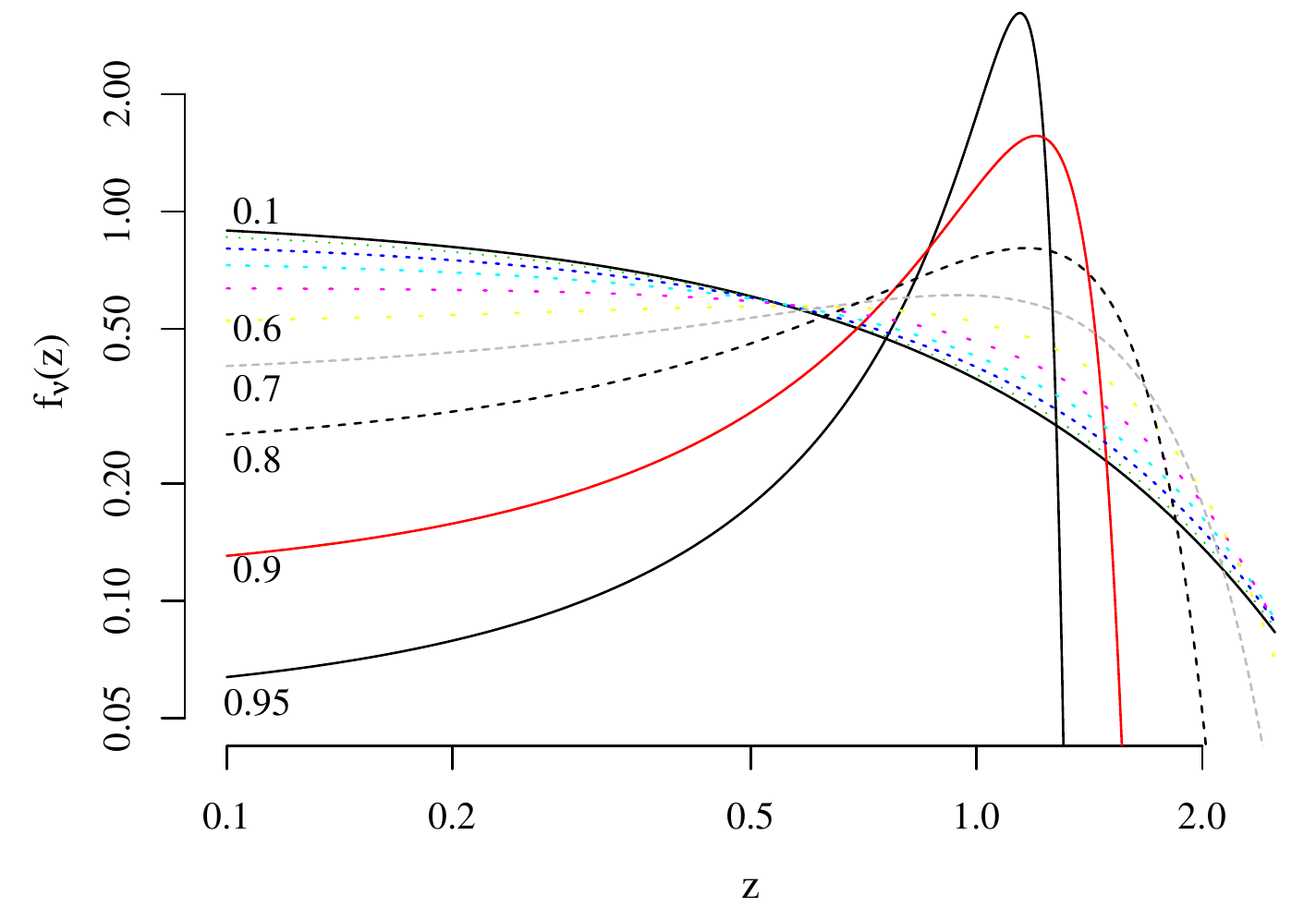}
\caption{Limiting distributions for $\nu =
0.1(0.1)0.9$ and 0.95.}
\end{figure}


\section{Method of Moments}
In this section we derive method-of-moments estimators for parameters $\nu$, and
$\mu$, based on the first two moments of a transformed random
variable $T$. It is important to emphasize that the Hill (1975), Pickands (1975), and Haan and Resnick (1980) estimators can be used to estimate these parameters as well. However, the above estimators are only  using a portion of the information contained in the data  making them statistically less efficient. It is this drawback that motivates us to look for estimators that utilize, or even optimize the use, of all the available  information in the data.

Recall that
\begin{equation}
T \stackrel{d}{=} \frac{|\ln U|^{1/\nu}}{\mu^{1/\nu}}S(\nu),
\label{be1}
\end{equation}
where $U$ has $U(0,\;1)$ distribution,  $S(\nu)$ is one-sided $\alpha-$stable, and the random variables $U$ and $S(\nu)$
are statistically independent. Since the first moment doesn't exist, we
consider the log-transformation   of the
original random variable $T>0$.

The above  formulation~(\ref{be1}) implies that
\begin{equation}
\ln(T) \stackrel{d}{=} \ln\left(\frac{|\ln
U|^{1/\nu}}{\mu^{1/\nu}}S(\nu)\right). \label{be2}
\end{equation}
Simplifying~(\ref{be2}), we get the equivalent expression
\begin{equation}
\ln(T) \stackrel{d}{=} \frac{1}{\nu} \ln\left(\frac{|\ln
U|}{\mu}\right)+ \ln(S(\nu)). \label{be3}
\end{equation}
Taking the expectation of (\ref{be3}), we obtain the equality
\begin{equation}
\textsf{E}\ln(T) =\frac{1}{\nu} \big[ \textsf{E} \ln(|\ln U|)-
\ln(\mu) \big]+ \textsf{E}\ln(S(\nu)). \label{be4}
\end{equation}
Our task now is to obtain the first moments of the random variables
$\ln(|\ln U|)$ and $\ln(S(\nu))$. We start by finding
the distribution of the first. Let $Y =|\ln U|
=-\ln U $. The random variable $Y$ has  the distribution  $e^{-y},\;y>0$. After  the monotone transformation
$X=\ln Y$, one easily shows that  $X$ has the probability density
function
\[ f_X(x) = e^{x-e^x},\qquad x \in
\mathbb{R}. \] Thus, the first moment of $\ln(|\ln U|)$ can now be calculated   as follows:
\begin{equation} \textsf{E} X =
\int_\mathbb{R}
xe^{x-e^x}dx=\int_{\mathbb{R}^+}\ln(y)e^{-y}dy=-\mathbb{C},\label{be5}
\end{equation}
where  $\mathbb{C} \cong 0.57721566490153286$ is the Euler's constant, see, e.g., Boros and Moll (2004).

The next step is to find the expectation of $\ln(S(\nu))$. Zolotarev (1986), p.~213-220,  shows that
\begin{equation}
\textsf{E}\ln(S(\nu)) = \mathbb{C}\left( \frac{1}{\nu}-1\right).
\label{be6}
\end{equation}
When (\ref{be5}) and (\ref{be6}) are substituted into (\ref{be4}),
the latter equality becomes
\begin{equation}
\textsf{E}\ln(T) =\frac{1}{\nu} \left( (-\mathbb{C})- \ln(\mu)
\right)+\mathbb{C}\left(
\frac{1}{\nu}-1\right)=-\frac{\ln(\mu)}{\nu} -\mathbb{C}.\label{be7}
\end{equation}
 From equation (\ref{be7}), we obtain
\begin{equation}
\mu = \text{exp}( -\nu[ \textsf{E}\ln(T) + \mathbb{C}] ).\label{be8}
\end{equation}

Alternatively, the second moment of the log-transformed random
variable $T$ is given by
\begin{align}
\textsf{E} \left[\ln(T)\right]^2&=\textsf{E} \left[\ln \left(\left(
\frac{|\ln
U|}{\mu}\right)^{1/\nu}S(\nu)\right)\right]^2 \notag\\
&  =\textsf{E} \left[\frac{1}{\nu}\ln \left( \frac{|\ln
U|}{\mu}\right) + \ln (S(\nu))\right]^2. \label{be9}
\end{align}
Expanding the right-hand side (RHS) of (\ref{be9}), we obtain the equality

\begin{align}
\textsf{E} \left[\ln(T)\right]^2&=\textsf{E}
\bigg[\frac{1}{\nu^2}\left( \ln (|\ln U| ) - \ln(\mu) \right)^2
+\frac{2}{\nu}\ln\left(\frac{|\ln
U|}{\mu}\right)\ln(S(\nu))+ \ln(S(\nu))^2 \bigg]\notag \\
&=\textsf{E} \bigg[ \frac{1}{\nu^2}\left( \ln (|\ln U| ) - \ln(\mu)
\right)^2 +\frac{2}{\nu}\ln(|\ln
U|)\ln(S(\nu))\notag\\
& \qquad-\frac{2}{\nu} \ln(\mu)\ln(S(\nu))+
\ln(S(\nu))^2 \bigg] \label{be10} \\
&=\textsf{E} \bigg( \frac{1}{\nu^2} \Big\lbrace  \left[ \ln (|\ln U|
)\right]^2 - 2\ln(\mu)\ln (|\ln U|
) + \ln(\mu)^2\Big\rbrace \notag \\
& \qquad  +\frac{2}{\nu}\ln(|\ln U|)\ln(S(\nu))- \frac{2}{\nu}
\ln(\mu)\ln(S(\nu)) + \ln(S(\nu))^2 \bigg). \notag
\end{align}

From another integral formula involving the Euler constant, we
can easily obtain
\begin{equation}
\textsf{E} \left[  \ln (|\ln U| ) \right]^2 =\textsf{E} X^2 =
\int_\mathbb{R} x^2e^{x-e^x}dx=\int_{\mathbb{R}^+}\ln(y)^2e^{-y} dy=
\mathbb{C}^2 + \frac{\pi^2}{6}.\label{be11} 
\end{equation}
Note that $\pi^2/6=\zeta (2)$ is the value of the Riemann zeta
function at the point 2. Furthermore, Bening et el. (2004) reveals that
\begin{equation}
\textsf{E} \left[  \ln(S(\nu))
\right]^2=\left(\frac{1}{\nu}-1\right)^2 \mathbb{C}^2 +
\frac{\pi^2}{6}\left( \frac{1}{\nu^2}-1\right). \label{be12}
\end{equation}
Using equation~(\ref{be11}), equation~(\ref{be12}), and the statistical
independence of two random variables  $U$ and $S(\nu)$, equation~(\ref{be10}) becomes
\begin{equation}
\textsf{E} \left[\ln(T)\right]^2=\frac{\pi^2}{3\nu^2}+\frac{\left(
\ln(\mu) \right)^2}{\nu^2}+\mathbb{C}^2
-\frac{\pi^2}{6}+\frac{2\mathbb{C}\ln(\mu)}{\nu}.\label{be13}
\end{equation}
From (\ref{be8}),
\begin{equation}
\ln(\mu) =  -\nu[ \textsf{E}\ln(T) + \mathbb{C}]. \label{be14}
\end{equation}
Substituting (\ref{be14}) into (\ref{be13}) and simplifying the
resulting expression, we get
\[
\textsf{E} \left[\ln(T)\right]^2-
\left[\textsf{E}\ln(T)\right]^2+\frac{\pi^2}{6}=\frac{\pi^2}{3\nu^2}.
\]
This implies that
\[ \nu^2=\frac{\pi^2}{3\left(\sigma_{\ln T}^2
+\pi^2/6\right)}.
\]
Thus, the method-of-moments estimator for $\nu$ is
\begin{equation}
\widehat{\nu}=\frac{\pi}{\sqrt{3\left(\widehat{\sigma_{\ln
T}^2} +\pi^2/6\right)}} \label{be15} 
\end{equation}
and, similarly,  from (\ref{be8}),
\begin{equation}
\widehat{\mu}=\exp\bigg( -\widehat{\nu}\,\big(
\widehat{\textsf{E}\ln(T)} + \mathbb{C}\big) \bigg)=\exp\bigg(
-\widehat{\nu}\,\big( \widehat{\mu_{\ln T}} + \mathbb{C} \big)
\bigg) \label{be16}
\end{equation}
is an estimator for $\mu$.




\section{Asymptotic Normality of the Estimators $\hat \nu$ and $\Hat \mu$}
We will show asymptotic normality of the above  estimators for $\nu$ and $\mu$.
The discussion in Section 5 implies that
\[
\textsf{E}\ln (|\ln U| )=-\mathbb{C},\quad \text{and}\quad \textsf{E}
\left[  \ln (|\ln U| ) \right]^2=\mathbb{C}^2+\frac{\pi^2}{6}.
\]
A further calculation using {\it Mathematica} shows that
\[
\textsf{E}\left[  \ln (|\ln U| ) \right]^3 = -
\mathbb{C}^3-\frac{\mathbb{C}\pi^2}{2}-2\zeta (3)
\]
and
\[
\textsf{E}\left[  \ln (|\ln U| ) \right]^4 =
\mathbb{C}^2\left(\mathbb{C}^2+\pi^2\right) +
\frac{3\pi^4}{20}+8\mathbb{C}\zeta (3).
\]
Additionally, we have
\[
\textsf{E}\ln(S(\nu)) = \mathbb{C}\left( \frac{1}{\nu}-1\right),
\]
and
\[
\textsf{E} \left[  \ln(S(\nu))
\right]^2=\left(\frac{1}{\nu}-1\right)^2 \mathbb{C}^2 +
\frac{\pi^2}{6}\left( \frac{1}{\nu^2}-1\right).
\]
Reference \cite{zol86} provides the following formula for   higher log-moments
of $S(\nu)$:
\[
\textsf{E}\left(\ln |S(\nu)|\right)^k=\left(d^kw_\nu (s)/ds^k
\right)\big|_{s=0},
\]
where
\[
w_\nu (s)=\frac{\Gamma (1-s/ \nu)}{\Gamma (1-s)}.
\]

To calculate these moments, we need to find the power series
expansion of $ w_\nu (s)$. This turns out to be easier if we first
expand
\[
\ln w_\nu (s) = \ln \Gamma (1-s/ \nu) - \ln \Gamma (1-s)
\]
into a power series, see, Bening et al. (2004). Using the log-gamma expansion
\[
\ln \Gamma (1-\theta) = \mathbb{C} \theta + \sum
\limits_{k=2}^\infty \frac{ \zeta (k)}{k} \theta^k ,
\]
we get
\begin{align}
\ln w_\nu (s) &=\mathbb{C}\left( \frac{1}{\nu}-1\right)s +
\frac{\pi^2}{12} \left( \frac{1}{\nu^2}-1\right)s^2 +
\frac{1}{3}\zeta (3)\left(\frac{1}{\nu^3}-1 \right)s^3 \notag\\
& + \frac{1}{4}\zeta (4)\left(\frac{1}{\nu^4}-1 \right)s^4+
\frac{1}{5}\zeta (5)\left(\frac{1}{\nu^5}-1 \right)s^5 + O(s^6),
\notag
\end{align}
and, hence,
\begin{align}
w_\nu (s) &=1 + \mathbb{C}\left( \frac{1}{\nu}-1\right)s + \bigg[
\frac{\pi^2}{12} \left(
\frac{1}{\nu^2}-1\right)+\frac{1}{2}\mathbb{C}^2\left(
\frac{1}{\nu^2}-1\right)^2\bigg] s^2 \notag \\
& + \bigg[\frac{1}{3}\zeta (3)\left(\frac{1}{\nu^3}-1
\right)+\frac{1}{6}\mathbb{C}^3  \left( \frac{1}{\nu}-1\right)^3
+\mathbb{C}\left( \frac{1}{\nu}-1\right)\left(
\frac{1}{\nu^2}-1\right)\frac{\pi^2}{12} \bigg]s^3 \notag \\
& +\frac{1}{1440}\bigg[
\bigg(\frac{1}{\nu^3}-\frac{1}{\nu^4}\bigg)\bigg(
60\mathbb{C}^4(\nu-1)^3-60\mathbb{C}^2\pi^2(\nu-1)^2(1+\nu) \notag \\
& +\pi^4(\nu-3)(1+\nu)(3+\nu) +480\mathbb{C}(\nu^3-1)\zeta (3)\bigg)
\bigg]s^4 +O(s^5). \notag
\end{align}
The $k$th log-moment of $S(\nu)$ is simply the coefficient of the term
$s^k/k!$ in the above power series expansion (can also be obtained via
$\left(d^kw_\nu (s)/ds^k \right)\big|_{s=0}$). In particular, the third and fourth log-moments can be shown to be
\[
\textsf{E} \left[  \ln(S(\nu))
\right]^3=\frac{-2(\nu-1)^3\mathbb{C}^3+\mathbb{C}\pi^2
(\nu-1)^2(1+\nu)-4(\nu^3-1)\zeta(3)}{2\nu^3},
\]
and
\begin{align}
\textsf{E} \left[  \ln(S(\nu)) \right]^4 &= \frac{1}{60}\bigg[
\bigg(\frac{1}{\nu^3}-\frac{1}{\nu^4}\bigg)\bigg(
60\mathbb{C}^4(\nu-1)^3-60\mathbb{C}^2\pi^2(\nu-1)^2(1+\nu) \notag \\
&+\pi^4(\nu-3)(1+\nu)(3+\nu) +480\mathbb{C}(\nu^3-1)\zeta (3)\bigg)
\bigg], \notag
\end{align}
respectively. In addition, the above  derivations   show that
\[
\mu_{\ln T}=-\left( \frac{\ln(\mu)}{\nu}+\mathbb{C}\right)
\;\;\text{and}\;\;\sigma_{\ln T}^2=\frac{\pi^2}{3}\left(
\frac{1}{\nu^2} - \frac{1}{2}\right).
\]
The  second-, third-, and fourth-order moments of $\ln T$ are
\[
\textsf{E}\left(\ln T
\right)^2=\mathbb{C}^2-\frac{\pi^2\left(\nu^2-2
\right)}{6\nu^2}+\frac{\ln (\mu) \left[ 2\mathbb{C}\nu+\ln
(\mu)\right]}{\nu^2},
\]
\[
\textsf{E}\left(\ln T \right)^3=-\frac{[\mathbb{C}\nu + \ln (\mu)
][2\mathbb{C}^2\nu^2-\pi^2(\nu^2-2)+2 \ln (\mu)(2\mathbb{C}\nu + \ln
(\mu))]}{2\nu^3} -2\zeta (3),
\]
and
\begin{align}
\textsf{E}\left(\ln T \right)^4 &= \frac{1}{60\nu^4}\Bigg\lbrace
60\mathbb{C}^4\nu^4-60\mathbb{C}^2 \nu^2(\nu^2-2) + \pi^4(28-20\nu^2
+ \nu^4) \notag \\
& + 60 \ln (\mu)[2\mathbb{C}\nu+\ln
(\mu)]\bigg(2\mathbb{C}^2\nu^2-\pi^2(\nu^2-2)+2\mathbb{C}\nu\ln
(\mu) + [\ln (\mu)]^2 \bigg)\notag\\
& + 480\nu^3[\mathbb{C}\nu + \ln (\mu) ]\zeta (3)
\Bigg\rbrace,\notag
\end{align}
respectively. We now calculate  higher-order central moments of the random variable
$\ln T$. After a tedious algebraic manipulation, we get
\begin{align}
\mu_3&=\textsf{E}\left(\ln T- \mu_{\ln T} \right)^3
\notag\\
&=\textsf{E}\Bigg \lbrace \frac{1}{\nu} \ln\left(\frac{|\ln
U|}{\mu}\right)+ \ln(S(\nu)) - \Bigg[-\left(
\frac{\ln(\mu)}{\nu}+\mathbb{C}\right)
\Bigg]\Bigg \rbrace^3\notag \\
&= -2\zeta (3)\notag
\end{align}
and
\[
\mu_4=\textsf{E}\left(\ln T- \mu_{\ln T}
\right)^4=\frac{\pi^4(28-20\nu^2+\nu^4)}{60\nu^4}.
\]
If we let
\[
\overline{\ln T} = \frac{\sum \limits_{j=1}^n\ln T_j}{n} \quad
\text{and}\quad \widehat{\sigma_{\ln T}^2}= \frac{\sum
\limits_{j=1}^n \left(\ln T_j-\overline{\ln T}\right)^2}{n}
\]
then, the standard 2-D Central Limit Theorem implies, as $n \to \infty$,  the following convergence in distribution:
\[
\sqrt{n}\left(
  \begin{array}{c}
    \overline{\ln T}_n-\mu_{\ln T} \\
    \widehat{\sigma_{\ln T}^2} - \sigma_{\ln T}^2  \\
  \end{array}
\right) \stackrel{d}{\longrightarrow}  \textsl{\Large{N}} \left[
  \begin{array}{ccc}
    \left(
      \begin{array}{c}
        0 \\
        0 \\
      \end{array}
    \right)
   &, & \left(
       \begin{array}{cc}
         \sigma_{\ln T}^2 & \mu_3 \\
         \mu_3 & \mu_4-\sigma_{\ln T}^4 \\
       \end{array}
     \right)
    \\
  \end{array}
\right],
\]
where $N(\vec \mu  ,\Sigma)$ represents the 2-D normal  distribution with mean $\vec \mu$, and covariance matrix $\Sigma$, and  $\mu_3,\mu_4,$ and $ \sigma_{\ln T}^2$ are defined above.

 Now, to show the asymptotic normality of the estimators $\hat \nu $ and $\hat \mu$,  we will rely on Cramer's Theorem (see, e.g., Ferguson ( 1996), p. 45,   which we are stating  below without proof.

\noindent \newtheorem*{thm3}{Theorem (Cramer)}
\begin{thm3}
Let $\mathbf{g}$ be a mapping $\bf{g}$$:$ $\mathbb{R}^d\to\mathbb{R}^k$ such that  $\dot{\bf{g}}(\bf{x})$ is continuous in a neighborhood of $\mbox{\boldmath$\theta$} \in \mathbb{R}^d$. If $\bf{X}_n$ is a  sequence of d-dimensional random vectors such that $\sqrt{n}(\bf{X}_n-\boldmath{\theta})\stackrel{d}{\to}\bf{X}$, then
\[
\sqrt{n}\big(\bf{g}(\bf{X}_n)-\bf{g}(\bf{\theta})\big)\stackrel{d}{\to}\bf{\dot{g}}(\bf{\theta})\bf{X}.
\]
In particular, if  $\sqrt{n}(\bf{X}_n-\bf{\theta})\stackrel{d}{\to}N(\bf{0},\bf{\Sigma)}$ where $\bf{\Sigma}$ is a $d \times d$ covariance matrix, then
 \[
\sqrt{n}\big(\bf{g}(\bf{X}_n)-\bf{g}(\bf{\theta})\big)\stackrel{d}{\to}N(\bf{0},\;\bf{\dot{g}}(\bf{\theta})\bf{\Sigma}\bf{\dot{g}}(\bf{\theta})^T).
\]
\end{thm3}

Indeed, for $\sigma_{\ln T}^2>0$,   Cramer's Theorem shows that
\begin{align}
\sqrt{n}\left(\widehat{\nu}-\nu\right)&
\stackrel{d}{\longrightarrow} \textsl{N} \left[0, \;
\frac{18\pi^2}{\left(6\sigma_{\ln T}^2+\pi^2
\right)^3}\left(\mu_4-\sigma_{\ln T}^4 \right) \right] \notag \\
& \stackrel{d}{\longrightarrow} \textsl{N} \left[0, \;
\frac{18\pi^2\big(\frac{\pi^4\left(32-20\nu^2-\nu^4\right)}{90\nu^4}\big)}{\left(6\sigma_{\ln
T}^2+\pi^2 \right)^3}\right] \notag \\
& \stackrel{d}{\longrightarrow} \textsl{N} \left[0, \;
\frac{\pi^6\left( 32 -20\nu^2 -\nu^4\right)}{5\left(6\sigma_{\ln
T}^2+\pi^2 \right)^3\nu^4} \right] \notag \\
& \stackrel{d}{\longrightarrow} \textsl{N} \left[0, \;
\frac{\nu^2\left( 32-20\nu^2-\nu^4\right)}{40} \right],\notag
\end{align}
where the last line of the preceding simplification is obtained  by substituting
$\sigma_{\ln T}^2=\frac{\pi^2}{3}\left( \frac{1}{\nu^2} -
\frac{1}{2}\right).$

Similarly, the estimator $\widehat{\mu}$ can be rewritten as
\[
\widehat{\mu}=\exp \Big(- \widehat{\nu}\left(\widehat{\mu_{\ln
T}}+\mathbb{C}\right)\Big) =\exp \left(
-\frac{\pi}{\sqrt{3(\widehat{\sigma_{\ln
T}^2}+\pi^2/6)}}(\widehat{\mu_{\ln T}}+\mathbb{C}) \right).
\]
Let
\[
\bf{g}(\mu_{\ln T},\sigma_{\ln T}^2) = \exp \left(
-\frac{\pi}{\sqrt{3(\sigma_{\ln T}^2+\pi^2/6)}}(\mu_{\ln
T}+\mathbb{C}) \right).
\]
The gradient then becomes
\[
\dot{\bf{g}}(\mu_{\ln T},\sigma_{\ln T}^2)= \left(
                                         \begin{array}{c}
                                           \frac{-\sqrt{2}\pi}{\sqrt{\pi^2+6\sigma_{\ln T}^2}}\exp \bigg(
\frac{-\sqrt{2}\pi(\mu_{\ln T}+\mathbb{C})}{\sqrt{\pi^2+6\sigma_{\ln
T}^2} } \bigg) \\
                                         \frac{3\sqrt{2}\pi(\mu_{\ln T}+\mathbb{C})}{\left(\pi^2+6\sigma_{\ln T}^2\right)^{3/2}}\exp \bigg(\frac{-\sqrt{2}\pi(\mu_{\ln T}+\mathbb{C})}{\sqrt{\pi^2+6\sigma_{\ln
T}^2} } \bigg)   \\
                                         \end{array}
                                       \right).
\]
By Cramer's theorem,
\[
\sqrt{n}\big( \widehat{\mu}-\mu\big) \stackrel{d}{\longrightarrow}
\textsl{N} \left[0,\; \sigma_{a}^2 \right],
\]
where
\begin{align}
\sigma_{a}^2 &= \dot{\bf{g}}(\mu_{\ln T},\sigma_{\ln T}^2)^T \left(
       \begin{array}{cc}
         \sigma_{\ln T}^2 & \mu_3 \\
         \mu_3 & \mu_4-\sigma_{\ln T}^4 \\
       \end{array}
     \right)
     \dot{\bf{g}}(\mu_{\ln
T},\sigma_{\ln T}^2) \notag \\
&=\frac{\mu^2\bigg[20\pi^4(2-\nu^2)-3\pi^2(\nu^4+20\nu^2-32)(\ln
\mu)^2}{120\pi^2} \notag\\
&\hskip 6cm  -\frac{720\nu^3(\ln \mu)\zeta (3)\bigg]}{120\pi^2}.
\end{align}
Therefore, we have shown that our method-of-moments estimators are
asymptotically normal (asymptotically unbiased). We can now
approximate the $(1-\varepsilon)100\%$ confidence interval for $\mu$,
and $\nu$ as follows:
\[
\widehat{\mu} \pm
z_{\varepsilon/2}\sqrt{\frac{\widehat{\mu}^2\bigg[20\pi^4(2-\widehat{\nu}^2)-3\pi^2(\widehat{\nu}^4+20\widehat{\nu}^2-32)(\ln
\widehat{\mu})^2-720\widehat{\nu}^3(\ln \widehat{\mu})\zeta
(3)\bigg]}{120\pi^2n}},
\]
and
\[
\widehat{\nu} \pm z_{\varepsilon/2}\sqrt{\frac{\widehat{\nu}^2\left(
32-20\widehat{\nu}^2-\widehat{\nu}^4\right)}{40n}},
\]
  where the tail quantile $z_{\varepsilon/2}$ is defined by  the equality
$P(Z>z_{\varepsilon/2})=\varepsilon/2$, with $Z \stackrel{d}{=}N(0,1)$.


\section{Testing   MoM estimators on  simulated data}
In this section we computationally compare and test the MoM estimators for $\nu$ and $\mu$ obtained in Section 5 using the mean
absolute deviation(MAD) from the true values of our parameters, and the square root of the mean squared error(MSE), as our
criteria. Recall our method-of-moment estimators for the fractional
order $\nu$ (\ref{be15}) and the intensity rate $\mu$ (\ref{be16}):
\begin{equation}
\widehat{\nu}_{mm}=\frac{\pi}{\sqrt{3\left(\widehat{\sigma_{\ln
T}^2} +\pi^2/6\right)}} \notag 
\end{equation}
and
\begin{equation}
\widehat{\mu}_{mm}=\exp\bigg( -\widehat{\nu}\,\big(
\widehat{\textsf{E}\ln(T)} + \mathbb{C}\big) \bigg)=\exp\bigg(
-\widehat{\nu}\,\big( \widehat{\mu_{\ln T}} + \mathbb{C} \big)
\bigg). \notag 
\end{equation}

\subsection{Simulated fPp}

We generate $n=100$ samples of the fPp jump times with sample sizes
N=100, 1,000, and 10,000. We then calculate the estimates $\hat \nu$ and $\hat \mu$   for each of the $n$ samples,  and then average them to obtain the means $\overline{\widehat{\nu}}$ and $\overline{\widehat{\mu}}$. These values are shown in the  tables below  together with their  MAD and
$\sqrt{\text{MSE}}$. The   fPp data were simulated for four different pairs of values    of $\mu$'s and $\nu$'s. The tables   show that , for the sample sizes $N =10^4$, the relative fluctuations the estimates of $\nu$, and $\mu$,  are  all below $5 $ percent.   However, judging this performance one must remember that in many practical applications, such as network traffic data, the typical sample sizes $N$ are of the order of millions, or more.

Furthermore, Tables \ref{table2}-\ref{table5} strongly suggest that  our  method-of-moments estimators
are asymptotically unbiased; they did fairly well in our simulations and  could be regarded as reasonable starting values for other iterative estimation procedures.

\begin{table}[h!t!b!p!]
\caption{\emph{Mean estimates of and  dispersions from the true parameter  for a simulated fPp data with $\left(\nu,
\;\mu\right)=\left(0.9, \;10\right)$.}}\label{t3} \centerline {
\begin{tabular*}{6.5in}{@{\extracolsep{\fill}}|l||c@{\hspace{0.01in}}c@{\hspace{0.01in}}c|c@{\hspace{0.01in}}c@{\hspace{0.01in}}c|c@{\hspace{0.01in}}c@{\hspace{0.01in}}c|}
 \hline
   &  & N=100 &  & & N=1,000 & &  & N= 10,000  & \\
 & Mean & MAD& $\sqrt{\text{MSE}}$& Mean& MAD &$\sqrt{\text{MSE}}$& Mean & MAD & $\sqrt{\text{MSE}}$\\
  \hline \hline
  $\widehat{\nu}_{mm}$ & .9027  & .0449  &.0556 & .9008  & .0128 & .0141 & .9012  & .0045  & .0056\\
  $\widehat{\mu}_{mm}$ & 10.06  & 1.289   & 1.649  & 10.05  & .4130  & .5130 &10.05  &.1386 &.1683\\
  \hline
\end{tabular*}
}
  \label{table2}
\end{table}

\begin{table}[h!t!b!p!]
\caption{\emph{Mean estimates of and dispersions from the true parameter  for a simulated fPp data with $\left(\nu,
\;\mu\right)=\left(0.3, \;1\right)$.}}\label{t4} \centerline {
\begin{tabular*}{6.5in}{@{\extracolsep{\fill}}|l||c@{\hspace{0.01in}}c@{\hspace{0.01in}}c|c@{\hspace{0.01in}}c@{\hspace{0.01in}}c|c@{\hspace{0.01in}}c@{\hspace{0.01in}}c|}
\hline
   &  & N=100 &  & & N=1,000 & &  & N= 10,000  & \\
 & Mean & MAD& $\sqrt{\text{MSE}}$& Mean& MAD &$\sqrt{\text{MSE}}$ & Mean & MAD & $\sqrt{\text{MSE}}$\\
  \hline \hline
  $\widehat{\nu}_{mm}$ & .3048  & .0233  & .0279 &.3001  & .0059 & .0073 & .3004   & .0021  &.0025 \\
  $\widehat{\mu}_{mm}$ & 1.025  & .1403  & .1789 &1.009  & .0473 & .0616 & .9998 & .0137& .0179 \\
  \hline
\end{tabular*}
}
    \label{table3}
\end{table}

\begin{table}[h!t!b!p!]
\caption{\emph{Mean estimates of and dispersions from the true parameter  for a simulated fPp data with $\left(\nu,
\;\mu\right)=\left(0.2, \;100\right)$. }}\label{t5} \centerline {
\begin{tabular*}{6.5in}{@{\extracolsep{\fill}}|l||c@{\hspace{0.01in}}c@{\hspace{0.01in}}c|c@{\hspace{0.01in}}c@{\hspace{0.01in}}c|c@{\hspace{0.01in}}c@{\hspace{0.01in}}c|}
\hline
   &  & N=100 &  & & N=1,000 & &  & N= 10,000  & \\
 & Mean & MAD&$\sqrt{\text{MSE}}$& Mean& MAD &$\sqrt{\text{MSE}}$ & Mean & MAD & $\sqrt{\text{MSE}}$\\
  \hline \hline
  $\widehat{\nu}_{mm}$ & .2062  & .0159   & .0197  &.2008   & .0041  & .0054 &  .1999  & .0013 & .0017\\
 $\widehat{\mu}_{mm}$ & 127.9 & 47.87 & 70.94 &102.3& 10.13  & 13.42 & 100.2 & 3.599 & 4.519 \\
  \hline
\end{tabular*}
}
    \label{table4}
\end{table}
\begin{table}[h!t!b!p!]
\caption{\emph{Mean estimates of and dispersions from the true parameter for a simulated fPp data with $\left(\nu,
\;\mu\right)=\left(0.6, \;1000\right)$.}}
\label{t6} \centerline {
\begin{tabular*}{6.5in}{@{\extracolsep{\fill}}|l||c@{\hspace{0.01in}}c@{\hspace{0.01in}}c|c@{\hspace{0.01in}}c@{\hspace{0.01in}}c|c@{\hspace{0.01in}}c@{\hspace{0.01in}}c|}
\hline
   &  & N=100 &  & & N=1,000 & &  & N= 10,000  & \\
 & Mean & MAD& $\sqrt{\text{MSE}}$& Mean& MAD &$\sqrt{\text{MSE}}$ & Mean & MAD & $\sqrt{\text{MSE}}$\\
  \hline \hline
  $\widehat{\nu}_{mm}$ & .6023  & .0378    & .0462  &.5999    &  .0119 & .0141     & .5998   &.0034 &.0042 \\
  $\widehat{\mu}_{mm}$ &1226  & 531.7  & 758.8  &1019 & 143.0 & 189.0& 997.4  & 38.68 & 48.56\\
  \hline
\end{tabular*}
}
    \label{table5}
\end{table}

\newpage

In Section 6 we have also derived the asymptotic probability distributions of our estimators which give the following 95 $\%$-confidence intervals for different values of the parameters to be estimated. Not surprisingly, they turned out to be much tighter than the bootstrap ones which were calculated using the built-in function in R, see, e.g. DiCiccio and Efron (1996). They were also better centered around the true values of the parameters.
For the ``Average" column of the tables shown below  we simulated 100 sets of sample size   $N$ and averaged the lower and upper $95\%$ confidence bounds calculated from the expressions obtained in Section 6.  For the ``Bootstrap" column we simulated 100 bootstrap replicates using the basic nonparametric bootstrap CI procedure. To see the   asymptotic behavior of the confidence intervals for larger N, our tables are given for N= 10,000, 100,000, and 1,000,000.
\medskip
  \begin{table}[h!t!b!p!]
\caption{\emph{95\% CI's  for a simulated fPp data with $\left(\nu,
\;\mu\right)=\left(0.9, \;10\right)$.}}\label{t3}
 \begin{small}
 \centerline {
\begin{tabular*}{6.5in}{@{\extracolsep{\fill}}|l||c@{\hspace{0.01in}}c|c@{\hspace{0.01in}}c|c@{\hspace{0.01in}}c|}
 \hline
 &  \multicolumn{2}{c|}{N=10,000} &  \multicolumn{2}{c|}{N=100,000} & \multicolumn{2}{c|}{N=1,000,000}\\   
 & Average & Bootstrap & Average & Bootstrap& Average & Bootstrap\\
  \hline \hline
  $\nu$ & (.8896,  .9113) &  (.8824,  .9107) & (.8967, .9036)     &  (.8926,  .9004)  & (.8988, .9010)    & (.8987,  .9008) \\
  $\mu$  & (9.668, 10.31) &  (9.563, 10.19)  & (9.900, 10.11)     &  (9.849, 10.059)  & (9.965, 10.03) & (9.948, 10.01)   \\
  \hline
\end{tabular*}
}
\end{small}
        \label{table6}
\end{table}
\begin{table}[h!t!b!p!]
\caption{\emph{95\% CI's  for a simulated fPp data with $\left(\nu,
\;\mu\right)=\left(0.3, \;1\right)$.}}\label{t4}
  \begin{small}
 \centerline {
\begin{tabular*}{6.5in}{@{\extracolsep{\fill}}|l||c@{\hspace{0.01in}}c|c@{\hspace{0.01in}}c|c@{\hspace{0.01in}}c|}
 \hline
 &  \multicolumn{2}{c|}{N=10,000} &  \multicolumn{2}{c|}{N=100,000} & \multicolumn{2}{c|}{N=1,000,000}\\  
 & Average & Bootstrap & Average & Bootstrap& Average & Bootstrap\\
  \hline \hline
  $\nu$ &  (.2947, .3049) &  (.2945,  .3042)& (.2985, .3017)     &  (.2988,  .3020)  & (.2994, .3004)    & (.2994,  .3004)  \\
  $\mu$ & (.9657, 1.035) &  (.9880,  1.061) &  (.9886, 1.010)   & (.9970,  1.022)  & (.9964, 1.003) & (.9970, 1.005)   \\
  \hline
\end{tabular*}
}
 \end{small}
          \label{table7}
\end{table}
\begin{table}[h!t!b!p!]
\caption{\emph{95\% CI's  for a simulated fPp data with $\left(\nu,
\;\mu\right)=\left(0.2, \;100\right)$. }}\label{t5}
\begin{small}
 \centerline {
\begin{tabular*}{6.5in}{@{\extracolsep{\fill}}|l||c@{\hspace{0.01in}}c|c@{\hspace{0.01in}}c|c@{\hspace{0.01in}}c|}
 \hline
 &  \multicolumn{2}{c|}{N=10,000} &  \multicolumn{2}{c|}{N=100,000} & \multicolumn{2}{c|}{N=1,000,000}\\ 
 & Average & Bootstrap & Average & Bootstrap& Average & Bootstrap\\
  \hline \hline
  $\nu$ &  ( .1966, .2035)   &  (.1995,  .2052) & (.1988, .2010)     &  (.1984,  .2006)  & (.1997, .2003)    & (.1998,  .2006)  \\
  $\mu$  & (91.48, 108.9) &  (94.60, 112.9) &  (97.11, 102.6)   & (96.79, 102.2)   & (99.15, 100.9) & (99.6, 101.3)  \\
  \hline
\end{tabular*}
}
\end{small}
      \label{table8}
\end{table}
\begin{table}[h!t!b!p!]
\caption{\emph{95\% CI's for a simulated fPp data with $\left(\nu, \;\mu\right)=\left(0.6, \;1000\right)$.}}
\label{t6}
 \begin{small}
\centerline {
\begin{tabular*}{6.5in}{@{\extracolsep{\fill}}|l||c@{\hspace{0.01in}}c|c@{\hspace{0.01in}}c|c@{\hspace{0.01in}}c|}
 \hline
 &  \multicolumn{2}{c|}{N=10,000} &  \multicolumn{2}{c|}{N=100,000} & \multicolumn{2}{c|}{N=1,000,000}\\   
 & Average & Bootstrap & Average & Bootstrap& Average & Bootstrap\\
  \hline \hline
  $\nu$ &  (.5906, .6091) &   (.5847,  .6016)& (.5968, .6026)     &  (.5985,  .6031)   & (.5990, .6008)    & (.5985,  .6003)  \\
  $\mu$   &(892.2, 1111)   & (840.2, 1028)&  (962.9, 1031)   & (960, 1036)  & (988.4, 1010) & (982.5, 1002)\\
  \hline
 \end{tabular*}
}
 \end{small}
           \label{table9}
\end{table}
\medskip

\section{Concluding remarks}

Our analysis shows that, in comparison to the  standard Poisson process,  the fractional Poisson process offers more modeling flexibility and and ability to accommodate some clumping (burstiness) in the set of the jump points of their   sample path, see Fig. \ref{Figure1}.  We have also succeeded in computing the
limiting distributions of the scaled $n$th arrival   time for the  fPp as well as
 the limiting distribution of $Z=N(t)/ \textsf{E}\big[N(t)\big]$ for fPp. Lastly, we were able to find asymptotically normal estimators of the
parameters of the fractional Poisson process.  The role of $\alpha$-stable densities turned out to be  critical  in analyzing the theoretical and numerical properties of fPp.

A number of interesting issues remain to be investigated including an extensions of our model to the fractional order   $1<\nu<2$,  and  to fractional Poisson
fields. The nonstationary fPp models permitting  nonconstant intensity rates  would be also of obvious interest. To the best of our knowledge, the multiscaling property and long-range
dependence of fPp has not been investigated either. Application of the above theory  to model real physical phenomena, such as network traffic, particle streams, economic ``events",  is in progress.


\section*{Appendix A. $\alpha^+$ stable densities}

The $\alpha^+$-density, or one-sided alpha-stable distribution,
denoted by $g^{(\alpha)}(t)$ is determined by its Laplace
transform as follows, see, e.g. Samorodnitsky and Taqqu (1994), and Uchaikin and Zolotariev (1999):
$$
\{\textsf{L}g^{(\alpha)}(t)\}(\lambda)\equiv\widetilde{g}^{(\alpha)}(\lambda)\equiv\int\limits_0^\infty
g^{(\alpha)}(t)e^{-\lambda t}dt=e^{-\lambda^\alpha}.\eqno(A.1)
$$
It is equal to 0 on the negative halfline, including the origin,
positive on the positive halfline,  and satisfies the normalization
condition
$$
\int\limits_0^\infty g^{(\alpha)}(t)dt=1.
$$

\begin{figure}[h!t!b!p!]
\centering
\includegraphics[width=.75\textwidth]{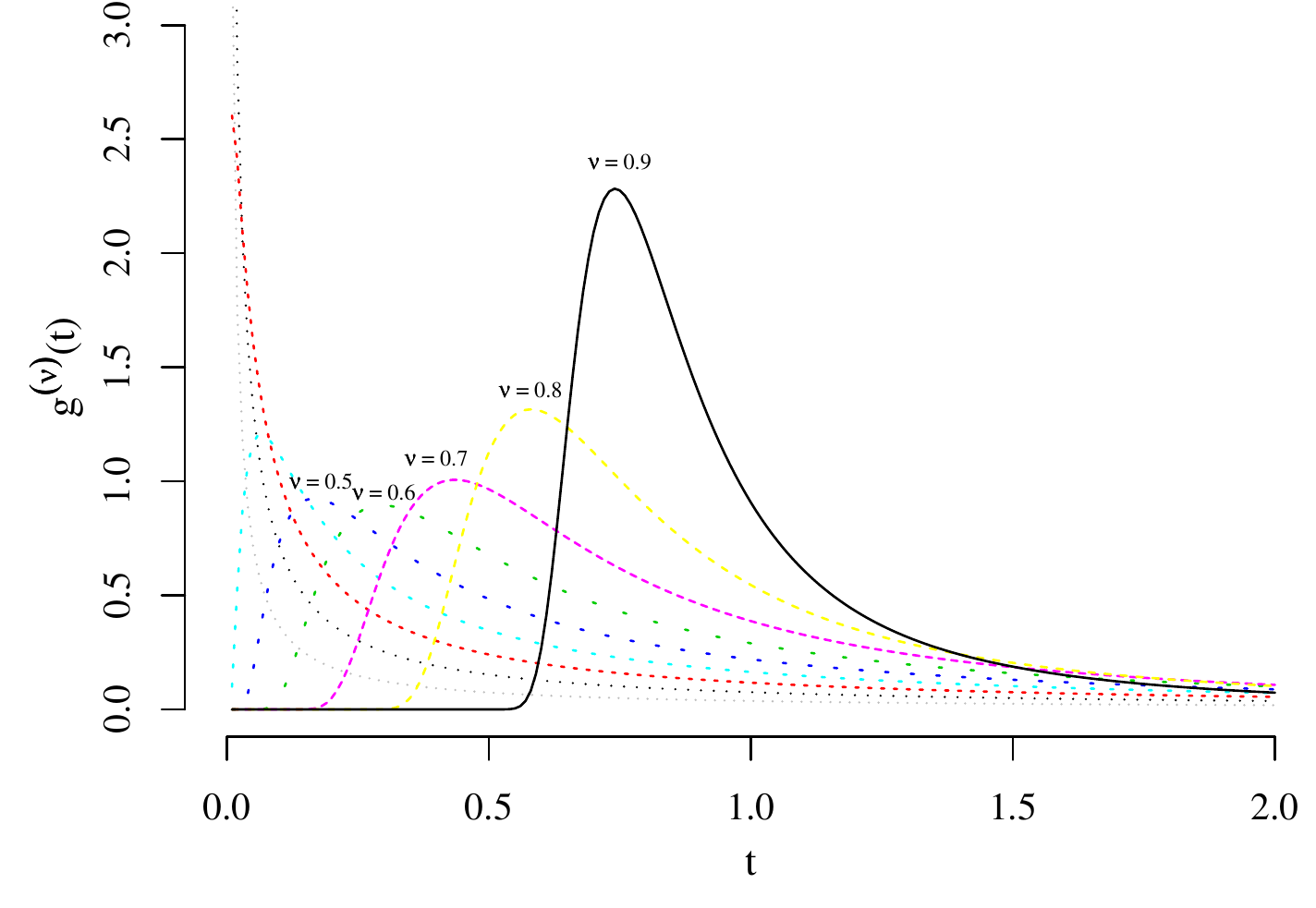}
\caption{\emph{$\alpha^+$-stable densities}}
\label{Figure 3}
\end{figure}

The term ``stable" means that these densities belong to the class
of the \textit{L'evy stable laws}: the convolution of two
$\alpha^+$-densities is again the $\alpha^+$-density (up to a
scale factor):
$$
\int\limits_0^tg^{(\alpha)}(t-t')g^{(\alpha)}(t')dt'=2^{-1/\alpha}g^{(\alpha)}(2^{-1/\alpha}t).
$$
This is easily seen in terms of Laplace transforms:
$$
\widetilde{g}^{(\alpha)}(\lambda)\widetilde{g}^{(\alpha)}(\lambda)=\widetilde{g}^{(\alpha)}(2^{1/\alpha}\lambda).
$$
Their role in the non-Gaussian   central limit theorem is crucial:  if
$T_1, T_2, \ldots, T_n$ are independent and identically
distributed random variables with with slowly decaying tail probabilities $P(T_j>t) \sim at^{-\alpha}$, $t
\to \infty$, then the probability density of their sum is, asymptotically, as $n \to \infty$,
$$
f_{\sum T_j}(t) \sim \left[ a\Gamma
(1-\alpha)\right]^{1/\alpha}g^{(\alpha)}\left(\left[a\Gamma
(1-\alpha)\right]^{1/\alpha}t\right)
$$

A few additional important properties of these densities are worth mentioning:

\medskip

(\textit{i}) If  $\alpha\to 1,$ then $\  g^{(\alpha)}(t)\to\delta(t-1);$

\medskip
(\textit{ii}) Moments of the $\alpha^+$ densities can be explicitly calculated:
$$ \int\limits_0^\infty
g^{(\alpha)}(t)t^{\nu}dt=\begin{cases}
\Gamma(1-\nu/\alpha)/\Gamma(1-\nu),& -\infty<\nu<\alpha;\\
\infty,& \nu\geq\alpha, \end{cases}\eqno(A.2)
$$

\medskip
(\textit{iii}) For $\alpha =1/2$ the density can be written out explicitly,
$$
g^{(1/2)}(t)= \frac{1}{2 \sqrt{\pi}} t^{-3/2} \exp[-1/(4t)],\
t>0,\eqno(A.3)
$$


(\textit{iv}) For numerical calculations, the following integral
formula is  convenient:
$$
g^{(\alpha)}(t) = \frac{\alpha t^{1/(\alpha-1)}}{\pi (1-\alpha)}
\int\limits_{-\pi/2}^{\pi/2} \exp \left\{ -t^{\alpha/(\alpha-1)}
U(\phi; \alpha)\right\} U(\phi; \alpha) d\phi,\eqno(A.4)
$$
where
$$
U(\phi; \alpha) = \left [ \frac{\sin (\alpha(\phi+\pi/2))}{\cos
\phi} \right]^{\alpha/({\alpha-1})} \frac{\cos \left(
(\alpha-1)\phi + \alpha \pi/2 \right)}{\cos \phi};
$$
\medskip
(\textit{v}) The following asymptotic  approximation may be obtained by the
saddle-point method:
$$
g^{(\alpha)}(t) \sim \frac{1}{\sqrt{2\pi(1-\alpha) \alpha}}
 (t/\alpha)^{(\alpha-2)/(2 - 2\alpha)}\exp [-(1-\alpha) (t/\alpha)^{-\alpha/(1-\alpha)} ], \quad t \to 0.
$$
Results of numerical calculations, using   (A.3), for
$\alpha=1/2$, and (A.6), for all other values of $\alpha$, are
shown  in Fig. {\ref{Figure 3}. For a complete discussion of $\alpha$-stable distributions, see the two monographs cited at the beginning of this Appendix.

\section*{Appendix B. Alternative fPp}
It is worth mentioning that there exists another  fractional generalization of the Poisson process   based on the analogy with the fractional Brownian
motion.  Instead of the stochastic differential equation
\[
\frac{d^\nu B_\nu}{dt^\nu}=W(t),
\]
where $W(t)$ is a Gaussian white noise, we can consider the equation
$$
\frac{d^\nu Y_\nu}{dt^\nu}=X(t),\eqno{(A.5)}
$$
where the random function $X(t)$ denotes the standard Poisson flow
\[
X(t) = \sum_{j=1}^{\infty} \delta(t-T^{(j)}),
\]
with $T^{(j)}=T_1 + T_2 + \ldots T_j$, and $ T_1, T_2,  \ldots T_j$
being  independent random variables with common density
\[
\psi (t)=\mu e^{-\mu t}, \qquad t \geq 0, \, \mu > 0.
\]
Integrating the stochastic fractional differential  equation (A.5) yields, see, e.g., Kilbas, Srivastava and Trujillo (2006),
$$
Y_\nu (t)  = \frac{1}{\Gamma (\nu)} \int _0^{t}
\frac{X(\tau)d\tau}{(t-\tau)^{1-\nu}} \notag
 = \frac{1}{\Gamma (\nu)}\sum^{N(t)}_{j=1} \int _0^{t}
\frac{\delta(\tau-T^{(j)})d\tau}{(t-\tau)^{1-\nu}}\notag
 = \sum^{N(t)}_{j=1}
\frac{1}{\Gamma (\nu)} \frac{1}{(t-T^{(j)})_+^{1-\nu}}.
$$
It is easy to see that, for $\nu=1$, the process becomes the standard
Poisson process. The stochastic process $Y_\nu$  can be interpreted as a  signal
generated by the Poisson flow of pulses, each of which giving the
contribution
$$
A (t-T^{(j)}) =\frac{1}{\Gamma (\nu)(t-T^{(j)})_{+}^{1-\nu}}.
\eqno{(A.6)}
$$
It is also well known that, conditional on  $N(t) = n $,   the unordered random  times $T^{(1)}, T^{(2)}, \ldots , T^{(n)}$ at
which events occur, are distributed independently and uniformly in the interval $(0,\;t)$.
Therefore,
\[
Y_\nu (t)|_{N(t)=n} = \sum^{n}_{j=1}A_j,
\]
where $A_j$ is determined by equation (A.6). Now,
\begin{align}
P(A_j > y) &= P\left(\Gamma (\nu) (t-T^{(j)})^{1-\nu} < y^{-1}
\right) \notag\\
&= P\left(t-T^{(j)} < \left[\Gamma (\nu)y\right]^{-1/(1-\nu)}
\right) \notag\\
&=P\left(T^{(j)} > t-\left[\Gamma (\nu)y\right]^{-1/(1-\nu)} \right)
\notag\\
& = P\left(T^{(j)} <\left[\Gamma
(\nu)y\right]^{-1/(1-\nu)}\right)
\notag\\
& = \frac{1}{t\left[\Gamma (\nu)y\right]^{1/(1-\nu)}}. \notag
\end{align}
Because $\nu > 0$, the expectation of $A_j$ exists, and according to the
law of large numbers, in this model the limit distribution of the scaled random
variable $Z$ (defined as in Section 4) has the degenerate limit distribution $f_\nu (z) =
\delta (z-1)$.  We will discuss statistical estimation procedures for this model in another paper.


\end{document}

%% file: abst2.tex
\begin{abstract}
The paper proposes an estimation procedure for parameters of the fractional Poisson
process (fPp) which is based on the method of moments (MoM). The basic tool is the
fractional calculus and the link between fractional Poisson process (fPp) and $\alpha$-stable
densities. Based on this result, we establish the asymptotic normality of our estimators
for the intensity rate $\mu$, and the fractional exponent $\nu$, two parameters appearing in
the fractional Poisson stochastic model; its properties are tested using synthetic data.

\emph{Keywords}: fractional Poisson process, $\alpha$-stable Lévy densities, fractional calculus,
asymptotic normality, method of moments estimators.

\end{abstract}